\documentclass[aps,twocolumn,nofootinbib,preprintnumbers,superscriptaddress]{revtex4-1}
\usepackage{amsmath,amssymb,amsfonts,amsthm,graphicx, epsf, dcolumn, yfonts}
\usepackage{latexsym}
\usepackage[colorlinks=true, pdfstartview=FitV, linkcolor=black, citecolor=black, urlcolor=black]{hyperref}
\usepackage{color}
\usepackage{slashed}
\newcommand{\be}{\begin{equation}}
\newcommand{\ee}{\end{equation}}
\newcommand{\bea}{\begin{eqnarray}}
\newcommand{\eea}{\end{eqnarray}}
\newcommand{\bwt}{\begin{widetext}}
\newcommand{\ewt}{\end{widetext}}
\newcommand{\wn}{\textswab{w}}

\def\del{\partial}

\newcommand{\alp}{\ensuremath{\alpha'}} 

\begin{document}

\title{Quantum Fluctuations in Holographic Theories with Hyperscaling Violation}

\author{Mohammad Edalati, Juan F. Pedraza and Walter Tangarife Garcia}
\affiliation{Theory Group, Department of Physics and Texas Cosmology Center, University of Texas at Austin, Austin, TX 78712, USA}

\preprint{UTTG-19-12}
\preprint{TCC-018-12}

\begin{abstract}

In this short note we use holographic methods to study the response of quantum critical points with hyperscaling violation to a disturbance caused by a massive charged particle. We give analytical expressions for the two-point functions of the fluctuations of the massive probe as a function of arbitrary (allowed) values of the hyperscaling violation exponent $\theta$ and the dynamical exponent $z$. We point out the existence of markedly different behaviors of the two-point functions in the parameter space of $\theta$ and $z$ at late times. In particular, as expected, the late-time dynamics of the probe becomes independent of its inertial mass in the range $z+2\theta/d>2$.

\end{abstract}

\maketitle

\section{Introduction}

Generalizing the notion of holography for gravitational theories in non-asymptotically AdS backgrounds is of great interest.
Such generalizations are not only interesting in their own right where one might hope to obtain a better understanding of quantum gravity in spacetimes other than AdS, but also from a practical point of view in terms of engineering toy models resembling the real-world non-relativistic  condensed matter systems. In the latter approach, the hope is that holography would shed light on some strongly-correlated features of these systems which would normally be impossible to understand using the conventional field theoretical methods.

Indeed, starting from the original works \cite{Son:2008ye,Balasubramanian:2008dm, Kachru:2008yh},  various holographic setups have already been constructed in the literature where the boundary theory is scale, but not conformally, invariant.  These boundary theories typically have either Schr\"{o}dinger or Lifshitz symmetries. A variant of such setups where the gravity solution is not only characterized by a dynamical exponent $z$ (as in the Lifshitz case) but also by the so-called hyperscaling violation exponent $\theta$, has recently gained some attention \cite{Gouteraux:2011ce,Ogawa:2011bz, Huijse:2011ef, Shaghoulian:2011aa, Dong:2012se,Copsey:2012gw,Kim:2012pd,Cadoni:2012ea,Alishahiha:2012qu,Bueno:2012sd,Sadeghi:2012vv,Sadeghi:2012ix,Dey:2012fi,Alishahiha:2012cm, Bhattacharya:2012zu,Ammon:2012je,Kiritsis:2012ta,Cadoni:2012uf,Perlmutter:2012he,Dey:2012tg,Singh:2012un,Kim:2012nb,Narayan:2012hk}; some earlier studies include \cite{Gubser:2009qt, Charmousis:2010zz,Singh:2010zs,Iizuka:2011hg}. The interest in these solutions partly stems from the observation that the entanglement entropy computed holographically using these gravity solutions exhibits a logarithmic violation of the area law in the boundary theory for some values of $\theta$ \cite{Ogawa:2011bz,Huijse:2011ef}.  Since entanglement entropy computed for theories with Fermi surfaces also shows a logarithmic violation \cite{Wolf:2006zzb,Gioev:2006zz,Swingle:2009bf,Swingle:2010yi,Zhang:2011}, these bulk solutions have been proposed as potential gravity duals of field theories with Fermi surfaces even though there are no explicit fermions in the bulk\footnote{See also \cite{Hartnoll:2012wm} for a discussion of the issues which plague identifying the gravity solutions with hyperscaling violation exponent as gravity duals for field theories with Fermi surfaces.}.

In this paper, we consider zero-temperature gravity solutions with hyperscaling violation parameter and assume that they represent in the boundary a class of quantum critical points characterized by two parameters, $z$ and $\theta$. Our objective here is to compute the response of these quantum critical points to a disturbance caused by coupling them to a massive charged particle (which is represented in the bulk by a long fundamental string).  We give analytical expressions for the two-point functions of the zero-temperature (quantum) fluctuations of the massive charged probe for arbitrary values of $z$  and $\theta$. This enables us to show the existence of a crossover in the late-time behavior of these two-point functions in the two-dimensional parameter space of $z$ and $\theta$. More concretely, in the range $z+2\theta/d>2$, where $d$ denotes the spatial dimension of the boundary theory, the two-point functions become independent of the mass of the probe at late times. We also verify the fluctuation-dissipation theorem for the quantum fluctuations of the probe. As a check, we show that our results for $\theta=0$ reduce to the ones in \cite{Tong:2012nf} for holographic quantum critical points with Lifshitz scaling. Moreover, our results also apply to the recently constructed holographic theories \cite{Kim:2012nb} which are supposed to represent in the boundary a class of quantum critical points with hyperscaling violation but with Schr\"{o}dinger symmetries. In addition, we study the zero-temperature fluctuations of the charged probe in quantum critical points dual to the Reissner-Nordstr\"{o}m AdS background and verify that the results agree with the late-time behavior of the two-point functions when the $z\to \infty$ limit is taken.

This paper is organized as follows. In the next section we briefly review some facts about the gravity solutions with hyperscaling violation exponent especially their regime of validity and the allowed values of $z$ and $\theta$ imposed by the null energy condition. In section \ref{Fluctuations} we present analytical results for the two-point functions of the zero-temperature fluctuation of the massive probe and analyze their late-time behavior as a function of $z$ and $\theta$, followed by the verification of the fluctuation-dissipation theorem in section \ref{FlucDissip}.  In section \ref{ZInfinity} we do the analysis for holographic quantum critical points with $z=\infty$ by taking the extremal Reissner-Nordstr\"{o}m AdS black hole as the background. We conclude with some remarks and open questions for future directions.

\section{Preliminaries}

Our starting point is the following $(d+2)$-dimensional line element
\begin{align}\label{bckgr2}
ds^2 &\equiv G_{\mu\nu}dx^\mu dx^\nu\nonumber\\
&=   \frac{1}{r^{2\theta/d}}\left(-r^{2z} dt^2 + \frac{dr^2}{r^2} + r^2 d\vec{x}^2\right).
\end{align}
where $d$ denotes the number of spatial dimensions and $z$ are $\theta$ are the dynamical critical  and the hyperscaling violation exponents, respectively. Such a metric could be obtained, for example, as a solution (in the IR) to the equations of motion coming from a system of Einstein-Maxwell-diatonic scalar with Lagrangian density given by \cite{Ogawa:2011bz,Huijse:2011ef}
\begin{align}\label{EMDAction}
\mathcal{L}= \frac{1}{2\kappa^2}\left[{\cal R}-L^2Z(\Phi)F^2-2(\partial\Phi)^2-\frac{V(\Phi)}{L^2}\right],
\end{align}
where
\begin{align}
Z(\Phi)=Z_0^2e^{\alpha\Phi}, \qquad V(\Phi)=-V_0^2e^{\delta \Phi}.
\end{align}
with $\alpha$ and $\delta$ being some constants determining $z$ and $\theta$. Also, the constants $Z_0$ and $V_0$ are related to the effective coupling of the gauge field and the cosmological constant, respectively. The solutions for the gauge field $A=A_t(r)dt$ and the dilatonic scalar $\Phi(r)$ will not play any significant role in our following discussions, hence, we will not write them here. The metric \eqref{bckgr2} is the most general one that is spatially homogeneous and covariant under the scale transformations
\be\label{scalings}
t \to \zeta^z t,\,\,\,\, \vec{x} \to \zeta \vec{x},\,\,\,\, r \to \zeta^{-1} r,\,\,\,\,ds \to \zeta^{\theta/d} ds\,.
\ee

Some comments on the allowed values of $z$ and $\theta$ are in order. On the gravity side, the null energy condition implies important consequences for theories that admit a consistent gravity dual \cite{Ogawa:2011bz, Dong:2012se}. These conditions can be summarized as\footnote{As we alluded to earlier, theories with some special values of $\theta$, namely for $\theta=d-1$, are of interest since they have been argued in \cite{Ogawa:2011bz,Huijse:2011ef} to give holographic realizations of theories with Fermi surfaces. The null energy condition then requires that the dynamical critical exponent satisfies $z \ge 2 - 1/d$ in order to have a consistent gravity description.}
\bea\label{NEC}
&&(d-\theta)\left[d(z-1)-\theta\right]\geq 0\,,\nonumber\\
&&(z-1)(d+z-\theta)\geq 0\,.
\eea
In a Lorentz invariant theory, $z=1$ and then the first inequality above implies that $\theta \le 0$ or $\theta \ge d$. On the other hand, for a scale invariant theory, $\theta=0$ and one recovers the known result $z \ge 1$ \cite{Kachru:2008yh, Koroteev}. Notice that, if $\theta\neq0$ the null energy condition  can be satisfied for $z<1$. In particular, $z<0$ together with $\theta>d$ gives a consistent solution to (\ref{NEC}), as well as $0<z<1$ along with $\theta \ge d+z$. However, as discussed in \cite{Dong:2012se}, $\theta > d$ leads to instabilities on the gravity side. Hence, we will not consider the case of $\theta > d$ here.

The metric \eqref{bckgr2}, together with the solutions for the gauge field and the dilatonic scalar,  is assumed to holographically describe a quantum field theory at a strongly-coupled quantum critical point with a dynamical critical exponent $z$ and a hyperscaling violation exponent $\theta$. As is well known in holography, the radial direction is mapped into the energy scale in the boundary field theory. For $\theta<d$, in the coordinates we have chosen in \eqref{bckgr2},  $r \to \infty$ and $r\to 0$ then describe, respectively, the UV  and IR of the field theory.  However, it is important to emphasize that the gravity background provides a good description of the aforementioned quantum critical point only in a certain range of $r$ as the solution could get significantly modified as the two regions $r \to \infty$ and $r \to 0$ are approached. If the dual field theory under consideration flows from a UV fixed point to a quantum critical point  which violates hyperscaling relation, then the background (\ref{bckgr2}) is only valid up to a scale of order $r\sim r_{\rm F}$ beyond which it ceases to exist as a valid solution to the equations of motion coming from the action \eqref{EMDAction} (Figure \ref{figHSV} depicts the regime of validity of our solution). The region $r\gtrsim r_{\rm F}$ in this case is drastically modified and the scale $r_{\rm F}$ then appears in the metric as an overall factor $ds^2 \propto L^2/r_{\rm F}^{2\theta/d}$ (with $L$ being the AdS radius) which is indeed responsible for restoring the canonical dimensions in the presence of hyperscaling violation\footnote{For example, in models with a Fermi surface, $r_{\rm F}$ is set by the Fermi momentum \cite{Ogawa:2011bz}. In addition, see \cite{Dong:2012se} for an example of UV completions of these models.}.  In the deep IR, on the other hand, the theory may flow to some other fixed points, develop a mass gap and so forth, resulting in the metric (\ref{bckgr2}) not being valid in this regime either \cite{Dong:2012se}. Relatedly, in the deep IR, the background seems to have a  genuine null singularity \cite{Shaghoulian:2011aa} for generic values of $z$ and $\theta$ allowed by the null energy condition, which may require stringy effects for it to be resolved. For now, we will simply ignore these issues while being cognizant of the fact that the  results we present in the following sections may only be valid in a certain range of energies.

\section{Quantum Fluctuations and\\ the two-point function}\label{Fluctuations}

A charged heavy particle on the boundary theory can be realized as the
endpoint of an open string that stretches between a D-brane and the IR region, $r = r_\epsilon \rightarrow0$, of the geometry. The brane is treated in the probe approximation. It covers the directions parallel to the boundary and spreads along the radial
direction from $r\rightarrow\infty$ to $r=r_b$ where it ends smoothly. For the validity of our computations, we will assume that $r_b\lesssim r_{\rm F}$, so that the above-mentioned issues regarding the validity of the gravity solution does not affect our analysis.

\begin{figure}
\begin{center}
\setlength{\unitlength}{1cm}
\includegraphics[width=6cm]{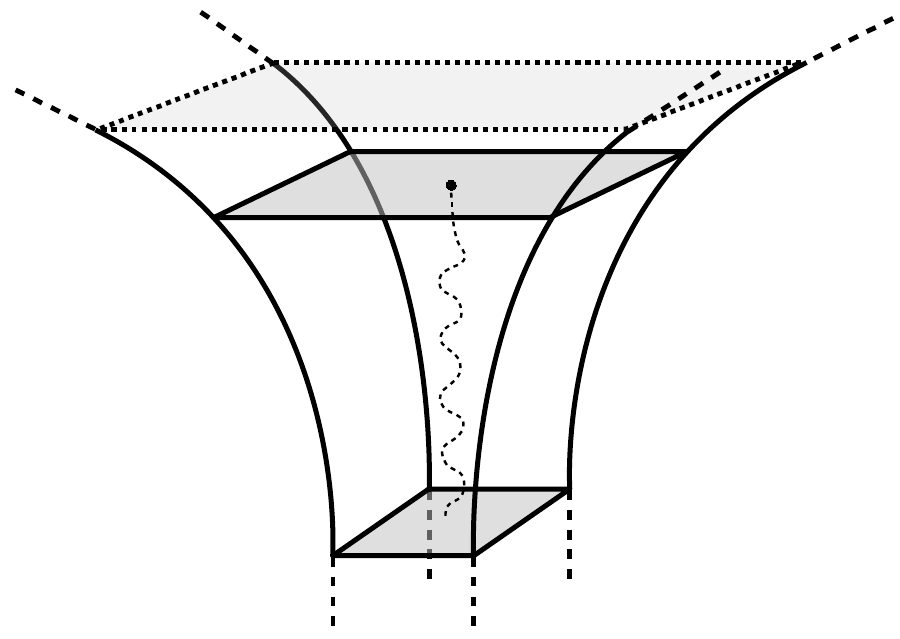}
 \begin{picture}(0,0)
 \put(0,4.05){\small{$r\to\infty$}}
  \put(-0.6,3.65){\small{$r=r_{\rm F}$}}
 \put(-1.45,3.05){\small{$r=r_b$}}
 \put(-2.15,0.85){\small{$r=r_\epsilon$}}
 \put(-2.15,0.4){\small{$r\to 0$}}
  \put(-4.4,0.2){\small{IR}}
  \put(-6.55,3.35){\small{UV}}
\end{picture}
\caption{Schematic picture of the fluctuating string dual to a charged heavy particle coupled to a quantum critical point. The string stretches in the bulk between a D-brane at $r=r_b$ and the IR region of the geometry at $r=r_\epsilon$.}
\label{figHSV}
\end{center}
\end{figure}
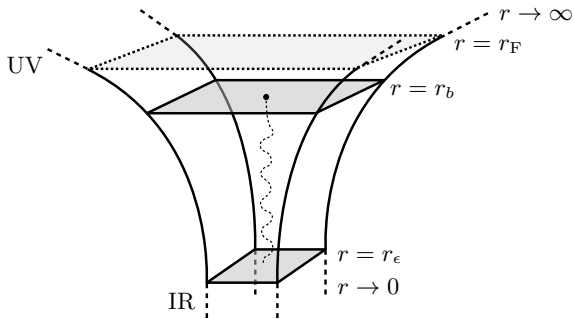
The classical dynamics of a fundamental string is governed by the Nambu-Goto action
\be\label{nambugoto}
S_{\text{NG}}=-\frac{1}{2\pi\alpha'}\int d^2\sigma\sqrt{-\det g_{ab}}\,,
\ee
where $g_{ab}=G_{\mu\nu}\del_aX^\mu\del_bX^\nu$ denotes the induced metric on the worldsheet. We choose to work in the static gauge, namely,  we set $\tau=t$ and $\sigma=r$. Our string embedding is then given by $X^\mu(t,r)=\{t,r,\vec{x}(t,r)\}$. One can then easily check that $\vec{x}(t,r)=\vec c$ is a trivial solution with $\vec c$ being a constant vector which we take, without loss of generality, to be zero. This solution is dual to a static particle whose energy is given by
\be \label{Edef}
E\,=\,\frac{1}{2\pi \alpha'}\int_0^{rb}dr\,\sqrt{-g_{tt} g_{rr}}\,=\,\frac{1}{2\pi  \alpha' }\frac{ r_b^{z-2 \theta/d}}{z-2\theta/d}.
\ee
It is worth mentioning that for $z\neq 1$, the energy $E$ is not the same as the inertial mass $m$ of the particle. The precise form of the dependence of $E$ on $m$ will be given below, after the computation of the response function.

Now, we would like to study the fluctuations over this static embedding (see Figure \ref{figHSV} for a schematic picture of the fluctuating string). Since the fluctuations along the various directions decouple from each other we restrict our attention to a single direction and take $X^\mu(t,r)=\{t,r,x(t,r),0,\cdots\}$.
Up to quadratic order in the perturbations $x(t,r)$, the Nambu-Goto action (\ref{nambugoto}) takes the form
\begin{align}\label{quadraticNGAction}
\hskip-0.1inS_{\text{NG}}\approx-\frac{1}{4\pi\alpha'}\int \hskip-0.03in dtdr\left(r^{z+3-2\theta/d}x'^2-r^{1-z-2\theta/d}\dot{x}^2\right),
\end{align}
where $\dot{x}\equiv\partial_t x$ and $x'\equiv\partial_r x$. Note that we dropped a constant term in \eqref{quadraticNGAction} that neither depends on $x$ nor on its derivatives. The resulting equation of motion is
\be\label{eom}
\frac{\partial}{\partial r}\left(r^{z+3-2\theta/d} \frac{\partial x}{\partial r}\right)-r^{1-z-2\theta/d}\frac{\partial^2 x}{\partial t^2}=0.
\ee
Now, because $t$ is an isometry of the background (\ref{bckgr2}), we Fourier transform $x(t,r)$:
\be\label{FTx}
x(t,r)\sim  e^{-i\omega t}g_\omega(r).
\ee
Substituting \eqref{FTx} into \eqref{eom}, the equation for the Fourier modes $g_\omega(r)$ becomes
\be\label{modeeom}
\frac{d}{dr}\left(r^{z+3-2\theta/d} \frac{d g_\omega}{dr}\right)+\omega^2r^{1-z-2\theta/d}g_\omega=0,
\ee
whose general solution can be written in the following form
\begin{align}\label{modsol}
g_\omega(r)= \frac{A_\omega}{r^{1+z/2-\theta/d}} & \left[J_{\frac{1}{2}+\frac{1}{z}-\frac{\theta}{zd}}\left(\frac{\omega}{z r^z}\right)+ \right. \nonumber\\
& \left. B_\omega Y_{\frac{1}{2}+\frac{1}{z}-\frac{\theta}{zd}}\left(\frac{\omega}{z r^z}\right)\right]\,.
\end{align}
Here $A_\omega,$ and $B_\omega$ are the two constants of integration and $J$ and $Y$ are Bessel functions of the first and second kind, respectively.

To fix the first constant we normalize the solution in the following way. For functions $u_1(t,r),u_2(t,r)$ satisfying the equation of motion (\ref{eom}), we can define a Klein-Gordon inner product \cite{Birrell:1982ix}
\be\label{innerp}
 \left(u_1,u_2\right)_\Sigma=-{i\over 2\pi\alp}\int_\Sigma \sqrt{\tilde g}\,
 n^\mu G_{xx} (u_1\partial_\mu u_2^*-\partial_\mu u_1\, u_2^*),
\ee
where $\Sigma$ is a Cauchy surface in the $(t,r)$ part of the metric, $\tilde g$ is the induced metric on
$\Sigma$ and $n^\mu$ is the future-pointing unit normal to $\Sigma$. This inner product is independent of the choice of $\Sigma$, but for simplicity we take $\Sigma$ to be a constant-$t$ surface.

We want a normalized basis of solutions such that for functions $u_\omega(t,r)=e^{-i\omega t}g_\omega(r)$, one has $(u_\alpha,u_\beta^*)=0$ and $(u_\alpha,u_\beta)=(u_\alpha^*,u_\beta^*)=\delta_{\alpha\beta}$. The reason is that, if this normalization is satisfied, it can be shown that the usual canonical commutation relations hold once the theory is quantized \cite{deBoer:2008gu}. In this context one would write
\begin{align}
x(t,r)=\int_{0}^\infty d\omega\, g_\omega(r) \left[a(\omega)e^{-i\omega t}+a^\dagger(\omega) e^{i\omega t}\right]\,, \label{xdef}
\end{align}
with $a^\dagger(\omega),a(\omega)$ being the creation and annihilation operators .

The first normalization condition is satisfied due to the properties of Bessel functions. The second one implies that,
\begin{align}
\left(u,u\right)&={1 \over 2\pi\alp}(\omega'+\omega)\int \frac{dr}{r^{z-1+\frac{2\theta}{d}}} g_\omega(r) g_{\omega'}(r)\nonumber\\
&=\delta(\omega-\omega').
\end{align}
Using (\ref{modsol}) and defining $\xi=1/r^z$, one obtains
\begin{align}
&{e^{-i(\omega-\omega') t} \over 2\pi\alp z} (\omega+\omega')  A_\omega A_\omega' \int d \xi\, \bigg\{ \xi\left[ J_\nu(\tfrac{\omega}{z}\xi)+B_\omega Y_\nu(\tfrac{\omega}{z}\xi)\right] \nonumber\\ &  \qquad\times \left[J_\nu(\tfrac{\omega'}{z}\xi)+B_{\omega'} Y_\nu(\tfrac{\omega'}{z}\xi)\right] \bigg\} =\delta(\omega-\omega'),
\end{align}
with $\nu=1/2+1/z-\theta/zd$. The last integral can be performed using various properties of Bessel functions\footnote{In particular, one needs to use the following properties
\begin{align}
&\int d\xi \,\xi J_\nu(u\xi)Y_\nu(v\xi)=0,\nonumber\\
&\int d\xi \,\xi J_\nu(u\xi)J_\nu(v\xi)=\int d\xi\, \xi Y_\nu(u\xi)Y_\nu(v\xi)=\frac{1}{u}\delta(u-v).\nonumber
\end{align}}. Finally, using the identity $\delta(ax)=\delta(x)/a$, we arrive at
\begin{align}\label{consta}
A_\omega=\sqrt{\frac{\pi\alp}{z(1+B_\omega^2)}}\,.
\end{align}
Note the particular dependence of $A_\omega$ on $B_\omega$, which itself remains to be fixed by the UV boundary condition. This is in stark contrast with the finite temperature case, in which case the overall normalization turns out to be sensible only to the IR part of the geometry \cite{Atmaja:2010uu,Fischler:2012ff}.

To fix the constant $B_\omega$, we impose Neumann boundary condition at the cut-off surface \cite{deBoer:2008gu}, \emph{i.e.} $x'(r_b)=0$.
A straightforward computation then yields
\be\label{constb}
B_\omega=- \frac{ J_{\frac{1}{z} - \frac{1}{2}-\frac{\theta}{zd}} \left( \frac{\omega}{z r_b^z} \right) } { Y_{\frac{1}{z} - \frac{1}{2}-\frac{\theta}{zd}} \left( \frac \omega {z r_b^z} \right) }.
\ee
Thus, the solution \eqref{modsol} takes the form
\bea
g_\omega(r)=\sqrt{\frac{\pi\alp}{z(1+B_\omega^2)}}&&\frac{1}{r^{1+z/2-\theta/d}}\left[J_{\frac{1}{2}+\frac{1}{z}-\frac{\theta}{zd}}\left(\frac{\omega}{z r^z}\right) \right. \nonumber \\ && \left. +B_\omega Y_{\frac{1}{2}+\frac{1}{z}-\frac{\theta}{zd}}\left(\frac{\omega}{z r^z}\right)\right]\,, \label{modsol2}
\eea
with the constant $B_\omega$ given in \eqref{constb}.

To calculate the two-point function, we use canonical commutation relations for the creation and annihilation operators, $[a(\omega),a^\dagger(\omega')]=\delta(\omega-\omega')$, and define the vacuum state such that $a(\omega)|0\rangle=0$. Then,
using equations (\ref{xdef}) and (\ref{modsol2}),  we obtain\footnote{In order to write the correlators in terms of the boundary theory data, one must specify the relation between $r_b$ and the mass $m$ of the charged particle. This can be obtained through the computation of the response function and is given at the end of section \ref{FlucDissip}. Additionally, $\alp$ needs to be written in terms of  the 't Hooft coupling $\lambda$. However, this last step depends on the UV completion of the effective gravitational model we are considering here.
We leave our results for the correlators in terms of $\alp$, while keeping in mind that this relation is implicit.}
\begin{align}
\label{twoptfn1}
\langle X(t)X(0) \rangle= \int_0^\infty \frac{d \omega}{2  \pi}\,e^{-i\omega t} \langle X(\omega)X(0) \rangle\,,
\end{align}
where
\begin{align} \label{twoptfn2}
&\langle X(\omega) X(0) \rangle\ = \frac{8 z \alp r_b^{-2+z+2 \theta /d}}{\omega^2}\nonumber \\
&\times \left[ J_{\frac{1}{z} - \frac{1}{2}-\frac{\theta}{zd}} \left( \frac{\omega}{z r_b^z} \right)^2  +Y_{\frac{1}{z} - \frac{1}{2}-\frac{\theta}{zd}} \left( \frac{\omega}{z r_b^z} \right)^2 \right]^{-1}.
\end{align}
The case of $z=1$, $\theta=0$ is the only one for which $\langle X(t) X(0) \rangle$ can be computed from $\langle X(\omega) X(0) \rangle$ analytically using \eqref{twoptfn1}.
For these special values of the parameters, the two-point function reads \cite{Tong:2012nf}
\be
\langle X(t) X(0)\rangle = -\frac 1 {4 \pi^2 \alpha 'E^2} \left(\log \vert t \vert + \gamma_E \right),
\ee
where $\gamma_E$ is the Euler-Mascheroni constant. For other values of the parameters an estimate for the behavior of $\langle X(t) X(0)\rangle$ at late times can be obtained as follows. At low frequencies, $\omega \ll r_b^z$, the leading order behavior of  $\langle X(\omega)X(0) \rangle$ reads
\be \label{twoptfhlw}
\hskip-0.1in\langle X(\omega)X(0) \rangle \sim \left\{
\begin{array}{l l}
\hskip-0.05inE^{\frac{2(z-2+2 \theta /d)}{z-2 \theta /d}} \omega^{-3+\frac{2}{z}-\frac{2 \theta }{z d}} & z+\frac{2\theta}{d}\leq2,\\
\hskip-0.05in\omega^{-1-\frac{2}{z}+\frac{2\theta }{z d}} & z+\frac{2\theta}{d}\geq2.
\end{array} \right. \hskip-0.1in
\ee
Assuming that at late times the dominant contribution to the two-point function comes from
the low frequency limit of $\langle X(\omega)X(0) \rangle$ as given in \eqref{twoptfhlw}, we find that
\be \label{twoptflt}
\hskip-0.1in\langle X(t)X(0) \rangle\sim \left\{
\begin{array}{l l}
\hskip-0.05inE^{\frac{2(z-2+2 \theta /d)}{z-2 \theta /d}} |t|^{2-\frac{2}{z}+\frac{2 \theta }{z d}} & z+\frac{2\theta}{d}\leq2,\\
\hskip-0.05in|t|^{\frac{2}{z}-\frac{2 \theta }{z d}} & z+\frac{2\theta}{d}\geq2.
\end{array} \right.\hskip-0.1in
\ee
Depending on the values of $z$ and $\theta$ the two-point function at late times shows markedly different behaviors.  In particular, notice that for $z+2\theta /d>2$, the long-time correlation of the particle is independent of the mass. (Note, in particular, that theories with $\theta=d-1$ belong to this category.) A similar change in behavior of the two-point function was recently shown in \cite{Tong:2012nf} for holographic theories with Lifshitz scaling but without hyperscaling violation.  Indeed, our results for $\theta=0$ perfectly agree with the analysis presented in \cite{Tong:2012nf}. Sitting exactly at the line $z+2\theta /d=2$, one can show that the two-point function grows linearly with $t$ at late times, which is its maximum rate of growth. The minimum, on the other hand,  can be realized in various situations: I) $\theta=d(1-z)$, II) $\theta=d$ with arbitrary $z$ or III) $z=\infty$ with arbitrary $\theta$, all of which give a logarithmic behavior in time for the late-time behavior of the two-point function.

The third situation deserves further attention. For any fixed $\theta$, taking the $z\to\infty$ limit, the second line in \eqref{twoptfhlw} implies that at low frequencies $\langle X(\omega)X(0) \rangle\sim \omega^{-1}$ regardless of the values of $\theta$ and $d$. Indeed , in section \ref{ZInfinity} we verify this behavior independently by considering the extremal Reissner-Nordstr\"{o}m AdS black hole. Since the near horizon geometry of an extremal  Reissner-Nordstr\"{o}m AdS black hole contains an AdS$_2$ factor, the boundary field theory flows in the IR to a quantum critical point with $z=\infty$ (which is holographically dual to AdS$_2$).

\section{Response function and the fluctuation-dissipation theorem}\label{FlucDissip}

We now turn to the computation of the response of the system due to an external force $F(t)$. From the point of view of the field theory, for $F(t)\sim e^{-i\omega t}F(\omega)$, the linear response of the particle is
\be
\left\langle x(\omega)\right\rangle=\chi(\omega)F(\omega),
\ee
where $\chi(\omega)$ is the retarded Green's function (also known as admittance). This can be easily realized from the gravity side by turning on a gauge field on the D-brane. Since the endpoint of the string is charged, this amounts to adding a minimal coupling to the action $S=S_{\text{NG}}+S_{\text{EM}}$, where
\be\label{actem}
S_{\text{EM}}=\int dt\left(A_t+\vec{A} \cdot\dot{\vec{x}}\right)\bigg|_{r=r_b}\,.
\ee
This will exert the desired force on the fluctuating particle. However, this coupling is just a boundary term, so it will not play any role for the dynamics of the string in the bulk. The UV boundary condition for the string is now replaced by
\be\label{NewUVbc}
\frac{\partial \mathcal{L_{\text{NG}}}}{\partial x'}\bigg|_{r=r_b}=-\frac{r_b^{3+z-2\theta/d}}{2\pi\alpha'}x'(r_b,t)=F(t)\,,
\ee
whereas in the IR region we impose ingoing boundary condition which is the appropriate one for the computation of the retarded Green's function $\chi(\omega)$ \cite{Son:2002sd}. In order to identify the desired combination of $J$'s and $Y$'s, notice that near $r\sim0$,
\be\label{tortoisez}
S\sim\int dt dr_{*}(x'^2-\dot{x}^2)\,,
\ee
where we have defined the `tortoise' coordinate $r_{*}=r^{-z}/z$, such that the $(t,r_{*})$ part of the metric is conformally flat. In \eqref{tortoisez}, prime denotes the derivative with respect to $r_{*}$. In this coordinate system, the equation of motion near the horizon $r_{*}\to \infty$ behaves just like the wave equation in flat space, with solutions given by
\bea
&& x^{(\text{out})}(t,r)\sim e^{-i\omega(t+r_{*})}\sim e^{-i\omega(t+r^{-z}/z)}, \nonumber \\&& x^{(\text{in})}(t,r)\sim e^{-i\omega(t-r_{*})}\sim e^{-i\omega(t-r^{-z}/z)}.
\eea
Thus, from the IR behavior of the Bessel functions, we select the first Hankel function $H^{(1)}=J+iY$ as the combination that satisfies  ingoing boundary condition at the horizon. Up to a constant, we then have
\be
x(t,r)=e^{-i\omega t}\frac{A_\omega}{r^{1+z/2-\theta/d}}H^{(1)}_{\frac{1}{2}+\frac{1}{z}-\frac{\theta}{zd}}\left(\frac{\omega}{z r^z}\right)\,.
\ee
Given the boundary condition (\ref{NewUVbc}) we get
\be
F(t)=e^{-i \omega t} \frac{r_b^{1-z/2-\theta /d}}{2\pi\alp} \omega A_\omega H^{(1)}_{-\frac{1}{2}+\frac{1}{z}-\frac{\theta}{zd}}\left(\frac{\omega}{z r_b^z}\right)\,,
\ee
from which we can read off the admittance,
\be\label{admitt}
\chi(\omega)=\frac{2 \pi \alp}{\omega r_b^{2-2 \theta/d}}\frac{H^{(1)}_{\frac{1}{2}+\frac{1}{z}-\frac{\theta}{zd}}\left(\frac{\omega}{z r_b^z}\right)}{H^{(1)}_{-\frac{1}{2}+\frac{1}{z}-\frac{\theta}{zd}}\left(\frac{\omega}{z r_b^z}\right)}\,.
\ee
It is straightforward to show that the fluctuation-dissipation theorem holds in the present setup at zero temperature.  In particular, this theorem relates the two-point function to the imaginary part of the admittance,
\be \label{flucdis}
\langle X(\omega) X(0)\rangle = 2\left[n_B(\omega)+1\right]\,{\rm Im}\,\chi(\omega)
\ee
where $n_B(\omega) = (e^{\beta \omega}-1)^{-1}$ is the Bose-Einstein distribution. Of course, at zero temperature (where $\beta\rightarrow\infty$) one ends up only with the last term in the above equation. On the other hand, from (\ref{admitt}) and using properties of Bessel functions it follows that
\begin{align}
\hskip -0.1in{\rm Im}\,\chi(\omega)&= \frac{4 z \alpha' r_b^{-2+z+2 \theta/d} }{\omega^2} \nonumber\\
& \hskip-0.1in\times \left[ J_{\frac{1}{z} - \frac{1}{2}-\frac{\theta}{zd}} \left( \frac{\omega}{z r_b^z} \right)^2 +Y_{\frac{1}{z} - \frac{1}{2}-\frac{\theta}{zd}} \left( \frac{\omega}{z r_b^z} \right)^2 \right]^{-1}\hspace{-5mm},
\end{align}
thus providing an explicit check of the fluctuation-dissipation theorem in the presence of hyperscaling violation in our holographic setup.

Finally, for low frequencies this response function can be written as
\be \label{admit2}
\chi(\omega) \sim \frac{1}{m\,(i\omega)^2 +
\gamma\,(-i\omega)^{1+2/z-2\theta/zd}\,+\,\cdots}\,,
\ee
where
\begin{align} \label{mass}
m&=\frac{{r_b}^{2-z-2 \theta /d}}{2-z-2\theta /d} , \nonumber \\
\gamma&=\frac{\pi\left[1-i \tan\left(\frac{\pi }{z}-\frac{\pi  \theta }{z d}\right)\right]}{(2 i z)^{2/z-2 \theta / zd}\mathrm{\Gamma}\left(\frac{1}{2}+\frac{1}{z}-\frac{\theta }{d z}\right)^2}.
\end{align}
The constants  $m$ and $\gamma$ are interpreted as the inertial mass and the the self-energy of the particle. For $z+2\theta/d>2$,  the self-energy dominates over the inertial mass at low frequencies, which is consistent with the change in the behavior of  the two point function found in the previous section. (Recall that for $z+2\theta/d>2$ the two-point function was independent of mass.) More explicitly, from (\ref{mass}) one observes that under the scale transformations given in (\ref{scalings}), $m$ transform as
\be
m\to\zeta^{z+2\theta/d-2}m,
\ee
implying that for $z+2\theta/d>2$, $m$ is an irrelevant coupling in the boundary theory, which should not affect the dynamics at low energies.

\section{Quantum Fluctuations in \\Holographic QCPs with $z=\infty$ }\label{ZInfinity}
Consider a ($d+2$)-dimensional Einstein-Maxwell system with negative cosmological constant $2\Lambda=-d(d+1)/L^2$
\be \label{act}
 S = {1 \over 2 \kappa^2} \int d^{d+2} x \,
 \sqrt{-g} \left[\mathcal{R} -2\Lambda - L^2F_{\mu\nu} F^{\mu\nu} \right].
\ee
 The $(d+2)$-dimensional Reissner-Nordstr\"om AdS black hole background (hereafter denoted by RN-AdS$_{d+2}$) is a solution \cite{Romans:1991nq,Chamblin:1999tk} to the Einstein-Maxwell equations of motion coming from the above action with the metric and gauge field given by
\begin{align} \label{bck1}
ds^2 &=  -{r^2 \over L^2}f(r) dt^2  + {L^2 \over r^2} {dr^2 \over f(r)}+ {r^2 \over L^2} d\vec x^2, \\
A&= \mu \Big(1- {r_0^{d-1} \over  r^{d-1}}\Big)dt,
\end{align}
where
\begin{align}\label{fMexpressions}
f(r)= 1 - {M \over r^{d+1}}+ { Q^2 \over r^{2d}},  \qquad M = r_0^{d+1} + {Q^2 \over r^{d-1}_0},
\end{align}
with $r_0$ being the radius of the horizon given by the largest positive root of $f(r)$.
The Hawking temperature of the black hole \eqref{bck1} is given by
\begin{align} \label{temp}
T = {(d+1) r_0 \over 4 \pi L^2} \left[1 - {(d-1) Q^2 \over (d+1) r_0^{2d}} \right].
\end{align}
The RN-AdS$_{d+2}$ background is assumed to holographically describe a ($d+1$)-dimensional boundary field theory at finite temperature $T$, given by \eqref{temp}, and finite chemical potential $\mu$ which is determined by the asymptotic ($r\to\infty$) value of the bulk gauge field $A_t(r)$. The chemical potential is related to the charge density $Q$ through
\begin{align} \label{chem}
\mu =  \sqrt{\frac{d}{2(d-1)}}{Q \over L^2 r_0^{d-1}}.
\end{align}

For the present computation, however, we are interested in the case where the boundary theory is at zero temperature, \emph {i.e.}  when the RN-AdS$_{d+2}$ is extremal. In this case, the near horizon geometry  becomes AdS$_2\times \mathrm{R}^{d}$.  The holographic interpretation is that the boundary theory flows in the IR to a dimensional CFT (dual to AdS$_2$) which describes a quantum critical point in which only the time coordinate scales, namely it is a QCP with $z=\infty$.
Nonetheless, similar to the theories with hyperscaling violation, the holographic description is not valid in the deep IR and should only be though of as an effective description up to some IR scale below which it flows to another fixed point. This could be be traced back to the fact that  the black hole has indeed a finite horizon area at zero temperature, suggesting a large ground state degeneracy. While keeping in mind the possible limitations of our results, we will not be concerned with such issues here.

Setting $T=0$ in \eqref{temp} yields
\be
Q^2=\frac{d+1}{d-1}r_0^{2d},
\ee
which upon being substituted in \eqref{fMexpressions}-\eqref{chem} results in
\begin{align}
f(r)& =1-\frac{2d}{d-1}\left(\frac{r_0}{r}\right)^{d+1}+\frac{d+1}{d-1}\left(\frac{r_0}{r}\right)^{2d},\\
M&=\frac{2d}{d-1}r_0^d,\\
\mu&=\sqrt{\frac{d(d+1)}{2}}\frac{r_0}{L^2(d-1)}.
\label{extremalf}
\end{align}

One can easily check that a static string is again a trivial solution to the equations of motion coming from the Nambu-Goto action. For a such a string, one has
\be \label{ene}
E=\frac{1}{2\pi \alp}\int_{r_0}^{r_b}dr\sqrt{-g_{tt} g_{rr}} \approx\frac{r_b}{2\pi  \alp },\,\,\,\text{for } \,\,\,r_b\gg r_0,
\ee
where $r_b$ is the radial location of the probe D-brane from which the string hangs.  Now, similar to our discussion in previous sections, we take an ansatz of the form $X^{\mu}=\{t,r,x(t,r),0,\cdots\}$ for the fluctuations around the static solution.
Up to quadratic order, the Nambu-Goto action becomes
\be
S_{\text{NG}}\approx-\frac{L^2}{4\pi\alp}\int dtdr\left[\frac{r^4f(r)}{L^4}x'^2-\frac{\dot{x}^2}{f(r)}\right],
\ee
where $\dot{x}\equiv\partial_t x$ and $x'\equiv\partial_r x$. The resulting equation of motion is then
\be\label{eom2}
\frac{\partial}{\partial r}\left(r^4f(r) \frac{\partial x}{\partial r}\right)-\frac{L^4}{f(r)}\frac{\partial^2 x}{\partial t^2}=0\,.
\ee
We now proceed by expanding $x(t,r)$ in Fourier modes, \emph{i.e.} $x(t,r)\sim  e^{-i\omega t}g_\omega(r)$. Equation \eqref{eom2} then yields
\be\label{modeeom2}
\frac{d}{d\rho}\left(\rho^4f(\rho) \frac{d g_\omega}{d\rho}\right)+\frac{\wn^2}{f(\rho)}g_\omega=0\,,
\ee
where, for convenience, we have defined dimensionless quantities
\be
\rho=\frac{r}{r_0}\,,\qquad \wn=\frac{L^2\omega}{r_0}.
\ee
Hereafter, primes will denote derivatives with respect to $\rho$ in our expressions. Also, we set $L=1$.

The next step is to find the solutions of the equation (\ref{modeeom2}).
For general $d$ one finds that, near the boundary, ${g_\wn}(\rho)$ has the following expansion
\be
g_\wn(\rho)={C}_1\Big(1+\frac{\wn^2}{2\rho^2}\Big)+{C}_2\frac{i\wn^3}{3 \rho^3}+\mathcal{O}\Big(\frac{1}{\rho^4}\Big). \label{sol1}
\ee
Note that $C_1$ and $C_2$ are functions of $\wn$.
To determine the constants of integration, $C_1$ and $C_2$, we have to study the behavior of (\ref{sol1}) in the IR, but since we are interested only in the solution at low energies, \emph{i.e.} $\wn\ll 1$, we perform a series expansion in $\wn$ and make use of a matching technique which can be found, for example, in \cite{deBoer:2008gu,Atmaja:2010uu,Fischler:2012ff}. For simplicity, and to reduce clutter in our expressions, we now focus on the case where $d=2$ (however, we expect our results to hold for general $d$, and  in the appendix we explicitly verify that this is indeed the case for $d=3$.)
Here, we only write down the final results,
relegating the details of the computations to the appendix.
For $d=2$ the constants of integration $C_1$ and $C_2$ take the form
\begin{align}\label{C1C2d2}
{C}^{(\text{out/in})}_1&=1\pm\frac{i \wn}{36} \left(\frac{\sqrt{2} }{2}\pi -\sqrt{2} \tan^{-1}\sqrt{2}-2\log 6\right),\nonumber \\
{C}^{(\text{out/in})}_2&=\mp\frac{1}{\wn^2}\,,
\end{align}
where the indices ``out'' and ``in'' correspond to outgoing and ingoing modes respectively.

To compute the two-point function $\langle X(\omega) X(0) \rangle$, we proceed differently compared to what we did in the previous sections. Namely, we first compute the response to an external force
\be
\langle X(\omega) \rangle \equiv \langle x(\omega, r_b) \rangle = \chi(\omega) F(\omega),
\ee
and then relate it to the two-point function, assuming that the fluctuation-dissipation theorem holds true
\be
\langle X(\omega) X(0) \rangle = 2 \,{\rm Im} \chi(\omega). \label{fluctdiss}
\ee
We impose ingoing boundary condition in the IR, namely $x(t,\rho)=A_\omega e^{-i \omega t}g^{(\text{in})}_\nu(\rho)$, where $A_\omega$ is an arbitrary constant that might have a frequency dependence. Using the UV boundary condition (evaluated at $\rho_b=r_b/r_0$)
\begin{equation}
\frac{r_0^3}{2\pi\alp}\rho^4f(\rho)x'\bigg |_{\rho_b}=F(t),
\end{equation}
and the solution (\ref{sol1}), we obtain
\begin{equation}
F(\omega)=\frac{r_b}{2\pi\alp} A_\omega f(\rho_b) \omega^2 \left({C}^{(\text{in})}_1+\frac{i \omega}{r_b}{C}^{(\text{in})}_2\right).
\end{equation}
In the limits $\omega/r_0 \ll 1$ and $r_b/r_0 \gg 1$, the imaginary part of the admittance scales in $\omega$ as
\be
\text{Im}\,\chi(\omega) \sim \frac{1}{\omega}.
\ee
Notice that this expression does not depend on the mass of the charged particle. (A similar result holds for  the case of $d=3$ as shown in the appendix.)
Consequently, at low frequencies, the scaling behavior of two-point function reads
\be
\langle X(\omega) X(0) \rangle \sim \frac{1}{\omega}.
\ee
Note that the correct dimensions in the above expression can trivially be restored by including powers of the chemical potential $\mu$. Some comments are in order here. First note that the low energy scaling of the two-function above, and the fact that the two-point function is also independent of mass, agrees with our previous results in the limit $z \rightarrow \infty$ (for arbitrary $\theta$). Such an agreement is not surprising given that geometries with Lifshitz scaling go over to AdS$_2\times R^d$ in the limit of $z\to\infty$.

\section{Final remarks}
In this note we studied the fluctuations of a heavy charged particle in a class of strongly-coupled
quantum critical points with dynamical exponent $z$ and hyperscaling violation exponent $\theta$. The late-time behavior of the two-point function for the zero-temperature fluctuations of the particle exhibits a crossover in the
$(z,\theta)$ parameter space. In a specific region, namely for $z+2\theta/d>2$, the two-point function is found to be independent of the mass.

Furthermore, we studied quantum critical points with $z=\infty$. Even though we focused on the cases $d=2$ and $d=3$, we expect that our results remain valid for arbitrary $d$. The reason is that the near horizon limit of RN-AdS$_{d+2}$ geometries have a universal behavior that goes over to AdS$_2\times \mathrm{R}^{d}$. This implies that at low energies the dual field theory exhibits emergent quantum critical behavior controlled by a CFT which could be though of representing a quantum critical point with $z=\infty$. Our results in this case are in agreement with the $z\to\infty$ limit of the behavior of the two-point function for the quantum fluctuation of a massive charged particle the in theories with hyperscaling violation, that, at late times, the two-point function grows logarithmically with time and is independent of the mass.

As a final remark, one might wonder if this markedly different behavior in the space of $z$ and $\theta$ holds true for other kind of operators. In \cite{Dong:2012se}, for instance, the authors considered massive scalars in the bulk and they found a transition in the two-point function from a universal power law at short distances (for $\theta>0$) to a nontrivial exponential behavior at long distances (where the WKB approximation is valid). It would be interesting to investigate this issue further in order identify more precisely the behavior in the full space of parameters $(z,\theta)$ and compare the results with our findings.

\section*{Acknowledgements}

This material is based upon work supported by the National Science Foundation under Grant No. PHY-0969020 and by the Texas Cosmology Center. W.T.G. is also supported by a University of Texas fellowship. We are grateful to D. Tong and K. Wong for correspondence.

\section*{Appendix}

In this appendix we will derive the solutions to the equation of motion (\ref{modeeom2}) given the boundary conditions to be discussed below. At low frequencies, the solutions can be obtained by means of a matching technique \cite{deBoer:2008gu,Atmaja:2010uu,Fischler:2012ff}. To find the solutions, consider three regimes: (I) the near horizon solution ($\rho\sim1$) for arbitrary $\wn$, (II) the solution for arbitrary $\rho$ in the limit $\wn\ll1$, and (III) the asymptotic $\rho\rightarrow\infty$ solution for arbitrary $\wn$. The idea is to find the approximate solutions for each of the three regimes and to match them to leading order in $\wn$. We implement the above matching method and write down the solutions only for the two cases of $d=2$ and $d=3$.

Before focusing on these two cases, let us make some remarks that are valid for arbitrary $d$. In terms of the `tortoise coordinate' defined by
\be\label{tort}
dr_{*}=\frac{dr}{r^2f(r)},
\ee
with the following behavior near the horizon
\begin{align}\label{tortoisehorizon}
r_{*}\sim\frac{1}{d(d+1)(r-r_0)}+\cdots\,,
\end{align}
we expect two solutions in the regime (I) of the form
\begin{align}
x^{(\text{out})}(t,r) &\sim e^{-i\omega(t+r_{*})}\sim  e^{-i\omega t}e^{-\frac{i\omega}{d(d+1)(r-r_0)}},\\ x^{(\text{in})}(t,r)& \sim e^{-i\omega(t-r_{*})}\sim e^{-i\omega t}e^{\frac{i\omega}{d(d+1)(r-r_0)}}\,,
\end{align}
corresponding to outgoing and ingoing modes,  respectively. The reason being is that, in this coordinate system, the $(t,r_{*})$ part of the metric is conformally flat and the equation of motion near $r\to r_0$ (or $r_{*}\to\infty$) behaves similar to the wave equation in flat space. In fact, near the horizon, the equation (\ref{modeeom2}) reduces to
\begin{align}
\left[d^2(d+1)^2(\rho -1)^2 g_\wn'(\rho )\right]'+\frac{\wn^2}{(\rho -1)^2}g_\wn(\rho)=0,
\end{align}
whose independent solutions are precisely given by\footnote{Note that the dots in (\ref{tortoisehorizon}) contain a subleading logarithmic divergence of the form $\sim D\log(\rho-1)$, for some constant $D$. This factor enters in the expressions for $g_\wn(\rho)$ as
\begin{align}
g_\wn^{(\text{out/in})}(\rho)&=(\rho-1)^{\mp i D\wn}e^{\mp\frac{i\wn}{d(d+1)(\rho-1)}} \nonumber \\
&= 1\mp \frac{i\wn}{d(d+1)(\rho-1)}\mp i\wn D \log(\rho-1)+\mathcal{O}(\wn^2)\,, \nonumber
\end{align}
but it does not affect the term of order $\mathcal{O}(1)$ in frequency.}
\begin{align}\label{solsA}
g_\wn^{(\text{out/in})}(\rho)&= e^{\mp\frac{i\wn}{d(d+1)(\rho-1)}}\nonumber\\
&=1\mp \frac{i\wn}{d(d+1)(\rho-1)}+\mathcal{O}(\wn^2)\,.
\end{align}

Asymptotically, one has $f(\rho)\to 1$, so the equation (\ref{modeeom2}) reduces to
\be\label{regC}
\frac{d}{d\rho}\left[\rho^4 g_\wn'(\rho )\right]+\wn^2 g_\wn(\rho)=0,
\ee
for arbitrary $d$. The general solution to (\ref{regC}) is given by
\begin{align}\label{regCinf}
g_\wn(\rho)&=A_1\left(1-\frac{i\wn}{\rho}\right)e^{i\wn/\rho}+A_2\left(1+\frac{i\wn}{\rho}\right)e^{-i\wn/\rho}\nonumber\\
&={C}_1\left(1+\frac{\wn^2}{2\rho^2}\right)+{C}_2\frac{i\wn^3}{3 \rho^3}+\mathcal{O}(1/\rho^4)\,,
\end{align}
where ${C}_1=A_1+A_2$ and ${C}_2=A_1-A_2$.

In the regime (II) one can can expand $g_\wn(\rho)$ as a power series in the frequency, \emph{i.e.}
\be\label{expnu}
g_\wn(\rho)=g^{(0)}_\wn(\rho)+\wn^2g^{(2)}_\wn(\rho)+\cdots\,.
\ee
The first term in the expansion satisfies the equation
\be\label{eomB}
\frac{d}{d\rho}\left[\rho^4f(\rho) \frac{d}{d\rho}g^{(0)}_\wn(\rho)\right]=0\,,
\ee
for which we have been able to find analytical solutions only for $d=2,3$. Therefore, we now turn our attention to these two particular cases.

\subsection{Solution for $d=2$}

The general solution of (\ref{eomB}) for $d=2$ reads
\begin{align} \label{solB3}
\hskip-0.07ing^{(0)}_\wn(\rho) = B_1 + B_2 &\left[  \frac{\sqrt{2} }{2}\tan^{-1}\left(\frac{\rho+1}{\sqrt{2}}\right) \right. \nonumber \\
 & \left.  + \log\left(\frac{3+\rho (\rho+2)}{(\rho-1)^2}\right)-\frac{3}{\rho-1}\right].
\end{align}
We can allow a frequency dependence for the constants of integration, but in order to have a reliable expansion as in (\ref{expnu}), we have to require that both $B_1$ and $B_2$ are at most linear in $\wn$. We now proceed to find these constants by expanding (\ref{solB3}) near the horizon and matching the solution with (\ref{solsA}). From (\ref{expnu}) and (\ref{solB3}) it follows that
\bea\label{solB3hor}
g_\wn(\rho)= B_1+B_2 && \left[\frac{\sqrt{2} }{2}\tan^{-1}\left(\sqrt{2}\right)+\log(6)-2\log(\rho-1) \right. \nonumber \\
&& \left. -\frac{3}{\rho-1}+\mathcal{O}(\rho-1)\right]+\mathcal{O}(\wn^2)\,.
\eea
Comparing the $\mathcal{O}(1/(\rho-1))$ and $\mathcal{O}(1)$ terms with the expression in (\ref{solsA}) we find that
\begin{align}\label{Bsoutin}
B^{(\text{out/in})}_1&=1\mp\frac{1}{18} i \wn \left(\frac{\sqrt{2} }{2}\tan^{-1}\sqrt{2}+\log 6\right), \\
B^{(\text{out/in})}_2&=\pm\frac{i \wn}{18}.
\end{align}
Finally, expanding the general solution in (\ref{solB3}) for $\rho\to\infty$ yields
\begin{align}\label{solB3inf}
\hskip-0.1ing_\wn(\rho)= B_1+B_2\left[\frac{\pi }{2 \sqrt{2}}-\frac{6}{\rho^3}+\mathcal{O}\Big(\frac{1}{\rho^4}\Big)\right]+\mathcal{O}(\wn^2).
\end{align}
Comparing equation \eqref{solB3inf} with (\ref{regCinf}) and using  (\ref{Bsoutin}), one obtains the expressions for $C_1$ and $C_2$ given in \eqref{C1C2d2}.

\subsection{Solution for $d=3$}

The general solution of (\ref{eomB}) for $d=3$ takes the form
\begin{align}\label{solB4}
g^{(0)}_\wn(\rho) = B_1 + B_2 &\Big[ 2 \sqrt{2} \tan^{-1}\Big(\frac{\rho}{\sqrt{2}}\Big)\nonumber\\
&+\tanh^{-1}(\rho )+\frac{3 \rho }{\rho ^2-1}\Big].
\end{align}
Again, to have a consistent expansion in frequencies, $B_1$ and $B_2$ are allowed to be at most of order $\mathcal{O}(\wn)$. From (\ref{expnu}) and (\ref{solB4}),  it follows that
\begin{align}\label{solB4hor}
&g_\wn(\rho)= B_1+B_2\left[\frac{3}{4}-\frac{\pi  i}{2}+\frac{1}{2}\log 2+2\sqrt{2}\cot^{-1}\sqrt{2} \right. \nonumber \\
& \left. -\frac{1}{2} \log(\rho-1)+\frac{3}{2 (\rho-1)}+\mathcal{O}(\rho-1)\right]+\mathcal{O}(\wn^2).
\end{align}
Comparing the $\mathcal{O}(1/(\rho-1))$ and $\mathcal{O}(1)$ terms with (\ref{solsA}), we obtain
\begin{align}\label{Bsoutin4}
B^{(\text{out/in})}_1&=1\pm\frac{i \wn}{36}\left(\frac{3}{2}-\pi i+4\sqrt{2}\cot^{-1}\sqrt{2}+\log 2\right),\\
B^{(\text{out/in})}_2&=\mp\frac{i\wn}{18}\,.
\end{align}
Expanding (\ref{solB4}) for $\rho\to\infty$ results in
\begin{align}\label{solB4inf}
g_\wn(\rho)&= B_1+B_2\left[\sqrt{2}\pi -\frac{\pi  i}{2}+\frac{6}{\rho^3}+\mathcal{O}\Big(\frac{1}{\rho^4}\Big)\right]\nonumber\\
&+\mathcal{O}(\wn^2)\,.
\end{align}
Comparing \eqref{solB4inf} with (\ref{regCinf}), we finally get
\begin{align}
{C}^{(\text{out/in})}_1&=1\pm\frac{i \wn}{36}\left(\frac{3}{2}-2\sqrt{2} \pi +4 \sqrt{2} \cot^{-1}\sqrt{2}+\log 2\right)\\
{C}^{(\text{out/in})}_2&=\mp\frac{1}{\wn^2}\,.
\end{align}
Going through the same steps as we did in section \ref{ZInfinity}, one easily obtains the following scaling behavior for the imaginary part of the admittance at low frequencies
\be
\text{Im}\,\chi(\omega) \sim\frac{1}{\omega},
\ee
which using the fluctuation-dissipation theorem results in
\be
\langle X(\omega) X(0) \rangle \sim \frac{1}{\omega}.
\ee

\end{document}